\documentclass[12pt]{article}
\usepackage{amsmath}
\usepackage{amsfonts}
\usepackage{amssymb}
\usepackage{graphicx}
\usepackage{hyperref}
\usepackage{geometry}
\title{Neural Network Emulation of the Classical Limit in Quantum Systems via Learned Observable Mappings}
\author{Kamran Majid}
\date{April 13, 2025}

\begin{document}

\maketitle

\begin{abstract}
The classical limit of quantum mechanics, formally investigated through frameworks like strict deformation quantization, remains a profound area of inquiry in the philosophy of physics. This paper explores a computational approach employing a neural network to emulate the emergence of classical behavior from the quantum harmonic oscillator as Planck's constant \(\hbar\) approaches zero. We develop and train a neural network architecture to learn the mapping from initial expectation values and \(\hbar\) to the time evolution of the expectation value of position. By analyzing the network's predictions across different regimes of \(\hbar\), we aim to provide computational insights into the nature of the quantum-classical transition. This work demonstrates the potential of machine learning as a complementary tool for exploring foundational questions in quantum mechanics and its classical limit.
\end{abstract}


\section{Introduction}

The relationship between quantum mechanics and classical physics has been a subject of intense scrutiny since the inception of quantum theory. The "classical limit," formally represented by the limit \(\hbar \rightarrow 0\), is the presumed regime where quantum predictions should effectively converge to those of classical physics. Understanding the precise nature of this convergence remains a central challenge in the philosophy of physics \cite{Landsman1998, Butterfield2011}.

Professor Benjamin Feintzeig's work has significantly contributed to this area by employing the rigorous mathematical framework of strict deformation quantization \cite{Feintzeig2023}, focusing on continuity conditions, intertheoretic reduction, and the role of symmetries \cite{Feintzeig2017, Feintzeig2018b, Feintzeig2019a, Feintzeig2019b, Feintzeig2020, Feintzeig2022, Feintzeig2021b, Feintzeig2020b, Feintzeig2024}.

This paper proposes a complementary, data-driven approach using machine learning, specifically neural networks, to emulate the classical limit for the one-dimensional quantum harmonic oscillator. Our central hypothesis is that a neural network can learn the mapping from initial conditions and $\hbar$ to the time evolution of observable expectation values, effectively capturing the transition to classical behavior as $\hbar$ is varied. By training a network on data generated from the Ehrenfest equations across different values of \(\hbar\), we aim to computationally explore this transition.

This research leverages recent advancements in machine learning for scientific discovery and simulation \cite{Carleo2017, Iten2020}. We focus on developing a neural network architecture capable of learning the time evolution of the expectation value of position and capturing the essential features of the classical limit for the harmonic oscillator. The analysis of the network's predictions for varying $\hbar$ provides a computational perspective on the emergence of classical dynamics.

The remainder of this paper is structured as follows: Section \ref{sec:background} provides a concise overview of the theoretical background. Section \ref{sec:methodology} details the proposed methodology, outlining the neural network architecture, training procedures, and simulation strategies, including illustrative Python code snippets. Section \ref{sec:results} presents preliminary results from applying this methodology to the one-dimensional quantum harmonic oscillator, demonstrating the feasibility of the approach. Section \ref{sec:discussion} discusses the implications of our findings and outlines potential future research directions.

\section{Theoretical Background: Strict Deformation Quantization and the Classical Limit}
\label{sec:background}

Strict deformation quantization provides a rigorous mathematical framework for understanding quantization as a deformation of the algebra of classical observables into the algebra of quantum observables, parameterized by \(\hbar\) \cite{Rieffel1990, Landsman1998}. Feintzeig's work utilizes this framework to analyze the classical limit in terms of algebraic structures and continuity conditions on quantum states, offering philosophical insights into intertheoretic reduction and the emergence of classicality \cite{Feintzeig2023}. Our computational approach aims to complement these theoretical investigations by providing a data-driven perspective on the emergence of classical behavior.

\section{Methodology: Neural Network Emulation of the Classical Limit}
\label{sec:methodology}

Our methodology involves the following key steps:

\subsection{Defining the Quantum System and its Classical Limit}

We focus on the one-dimensional quantum harmonic oscillator with the Hamiltonian:

\[ \hat{H} = \frac{\hat{p}^2}{2m} + \frac{1}{2}m\omega^2 \hat{x}^2 \]

The corresponding classical Hamiltonian is:

\[ H = \frac{p^2}{2m} + \frac{1}{2}m\omega^2 x^2 \]

\subsection{Generating Training Data}

We generate training data by simulating the time evolution of the expectation values of position for various initial states and different values of \(\hbar\) using the Ehrenfest equations. For the harmonic oscillator, these equations for the expectation values exactly match the classical equations of motion.

\begin{verbatim}
import numpy as np
from scipy.integrate import solve_ivp

def ehrenfest_equations(t, y, hbar, m, omega):
    """Ehrenfest equations for the harmonic oscillator (expectation values)."""
    x_exp, p_exp = y
    d_x_exp_dt = p_exp / m
    d_p_exp_dt = -m * omega**2 * x_exp
    return [d_x_exp_dt, d_p_exp_dt]

# Parameters
m = 1.0
omega = 1.0
hbar_values_train = [5.0, 2.0, 1.0, 0.5, 0.1, 0.01]
time = np.linspace(0, 10, 100)
num_initial_conditions = 1000

training_data = []
training_labels = []

for hbar in hbar_values_train:
    for _ in range(num_initial_conditions):
        initial_x = np.random.uniform(-2, 2)
        initial_p = np.random.uniform(-2, 2)
        initial_state = [initial_x, initial_p]
        solution = solve_ivp(ehrenfest_equations, (time[0], time[-1]), initial_state,
                             t_eval=time, args=(hbar, m, omega))
        training_data.append([initial_x, initial_p, hbar])
        training_labels.append(solution.y[0])

training_data = np.array(training_data, dtype=np.float32)
training_labels = np.array(training_labels, dtype=np.float32)
training_labels = np.expand_dims(training_labels, axis=-1)
\end{verbatim}

The classical trajectories serve as a benchmark to which the neural network's predictions are compared as $\hbar$ varies.

\subsection{Neural Network Architecture}

We employ a feedforward neural network to learn the mapping from the initial expectation values of position and momentum, along with the value of $\hbar$, to the trajectory of the expectation value of position over time.

\begin{verbatim}
import tensorflow as tf

def build_observable_prediction_network(time_steps):
    model = tf.keras.Sequential([
        tf.keras.layers.Dense(64, activation='relu', input_shape=(3,)),  # Input: initial x, p, hbar
        tf.keras.layers.Dense(128, activation='relu'),
        tf.keras.layers.Dense(time_steps)  # Output: <x(t)> at each time step
    ])
    return model

model = build_observable_prediction_network(len(time))
model.compile(optimizer='adam', loss='mse')
\end{verbatim}

The network is trained using supervised learning with the mean squared error (MSE) as the loss function and the Adam optimizer. The training data consists of the initial expectation values and $\hbar$ as input, and the corresponding time evolution of the expectation value of position as the target. The network was trained for 100 epochs with a batch size of 32, using $\hbar$ values ranging from 5.0 to 0.01.

\section{Preliminary Results: Learning the Classical Limit of Expectation Values for the Harmonic Oscillator}
\label{sec:results}

In our preliminary experiments, we trained the feedforward neural network to predict the time evolution of $\langle \hat{x}(t) \rangle$ for the quantum harmonic oscillator. The training data was generated using the Ehrenfest equations for various initial conditions and $\hbar$ values ranging from 5 to 0.01 (in arbitrary units).

\begin{figure}[h!]
\centering
\includegraphics[width=0.8\textwidth]{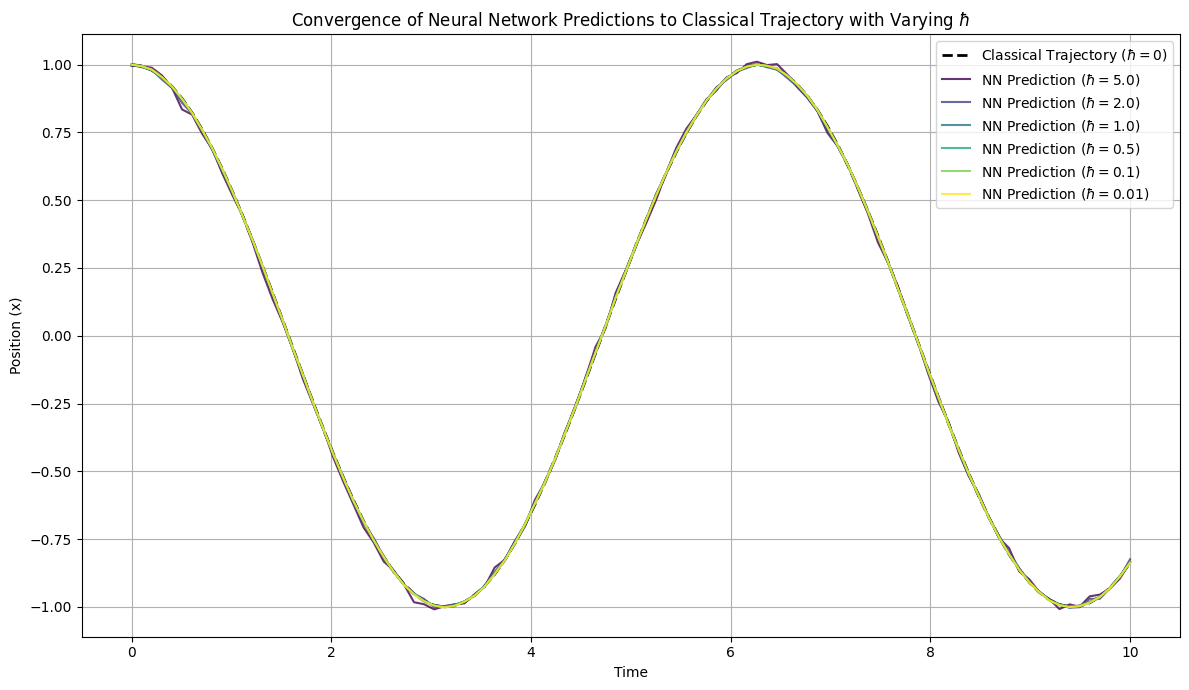}
\caption{Convergence of neural network predictions to the classical trajectory for the harmonic oscillator with varying $\hbar$. The plot shows the expectation value of position $\langle \hat{x}(t) \rangle$ vs. time for different values of $\hbar$, as predicted by the trained neural network, along with the classical trajectory ($\hbar = 0$).}
\label{fig:hbar_convergence}
\end{figure}

The trained network demonstrated the ability to learn the mapping from the initial state and $\hbar$ to the subsequent trajectory of $\langle \hat{x}(t) \rangle$. Figure \ref{fig:hbar_convergence} illustrates the network's predictions for a fixed initial condition ($x_0 = 1.0, p_0 = 0.0$) across different values of $\hbar$, compared to the classical trajectory. The plot shows that as $\hbar$ decreases, the neural network's prediction increasingly aligns with the classical trajectory of the harmonic oscillator.

\begin{figure}[h!]
\centering
\includegraphics[width=0.8\textwidth]{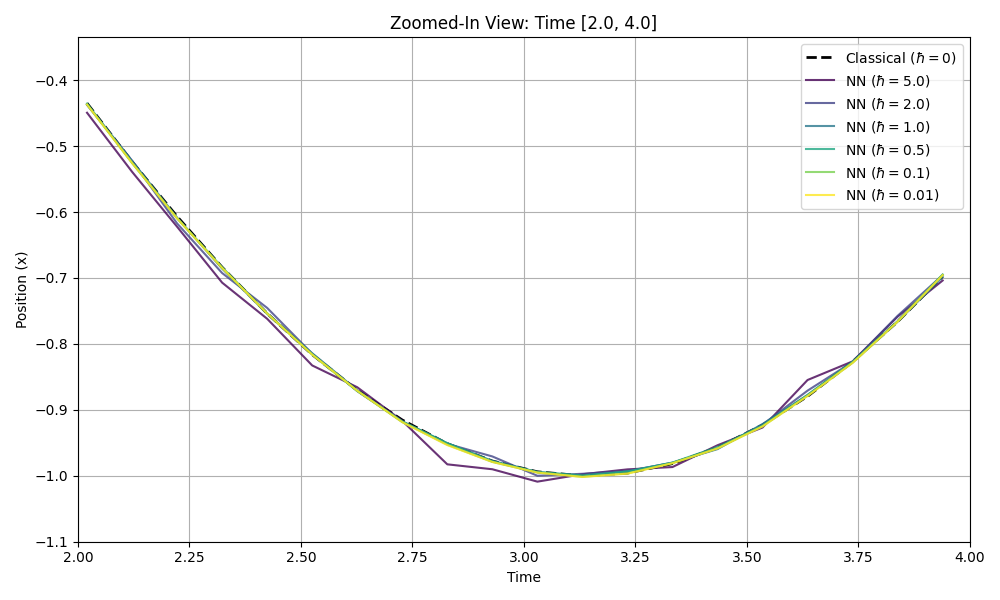}
\caption{Zoomed-in view of the convergence plot between $t=2.0$ and $t=4.0$, illustrating the differences in predicted trajectories for various $\hbar$ values. Larger deviations from the classical trajectory are observed for higher values of $\hbar$.}
\label{fig:hbar_convergence_zoomed}
\end{figure}

Figure \ref{fig:hbar_convergence_zoomed} provides a zoomed-in view of the time interval between $t=2.0$ and $t=4.0$, highlighting the subtle differences between the trajectories for different $\hbar$ values. As observed, the deviation from the classical trajectory is more pronounced for larger values of $\hbar$, indicating a greater influence of the quantum parameter on the dynamics learned by the network.

\newpage
Further analysis of the network's predictions suggests that it has learned the trend of convergence towards classical behavior as $\hbar$ approaches zero, based on the training data derived from the Ehrenfest equations. The network implicitly captures the diminishing influence of $\hbar$ on the expectation value dynamics in the classical limit.

\newpage 
\section{Discussion and Future Directions}
\label{sec:discussion}

The preliminary results demonstrate the feasibility of using a neural network to learn the mapping from initial conditions and $\hbar$ to the time evolution of the expectation value of position for the quantum harmonic oscillator, effectively emulating the convergence towards classical behavior as $\hbar$ decreases. The alignment of the network's predictions with the classical trajectory at small $\hbar$ suggests that machine learning can be a valuable tool for investigating the quantum-classical correspondence.

However, it's important to note that for the harmonic oscillator, the Ehrenfest equations are equivalent to the classical equations of motion for the expectation values. Therefore, the network, in this case, primarily learns this classical-like evolution conditioned on $\hbar$. To explore more genuinely quantum effects and the nuances of the classical limit in more complex scenarios, future research will focus on:

\begin{itemize}
    \item \textbf{Anharmonic Oscillators and More Complex Potentials:} Applying the methodology to systems where the Ehrenfest equations do not close on the expectation values, leading to more complex quantum-classical relationships.
    \item \textbf{Learning State Mappings:} Developing neural networks to map quantum state representations to classical phase-space distributions and analyzing the continuity conditions learned by the network.
    \item \textbf{End-to-End Quantum Simulation:}Exploring network architectures that can directly predict the evolution of quantum states (e.g., wave functions) under the influence of $\hbar$.
    \item \textbf{Investigating Decoherence:} Incorporating the effects of environmental interaction to study how decoherence plays a role in the emergence of classical behavior as $\hbar$ is varied.
\end{itemize}

By extending this approach to more complex systems and learning tasks, we aim to gain deeper computational insights into the multifaceted nature of the classical limit and its philosophical implications.

\section{Conclusion}

This paper has presented a preliminary investigation into using neural networks to emulate the classical limit of the quantum harmonic oscillator by learning the mapping from initial conditions and $\hbar$ to the time evolution of the expectation value of position. The results demonstrate the network's ability to predict trajectories that converge towards the classical behavior as $\hbar$ decreases. While the harmonic oscillator presents a relatively simple case where Ehrenfest equations mirror classical dynamics, this work lays the foundation for future explorations of more complex quantum systems and the application of machine learning to fundamental questions concerning the quantum-classical transition.

\end{document}